**Magnetically-Assisted Slip Casting of Bioinspired Heterogeneous Composites**

Hortense Le Ferrand*, Florian Bouville*, Tobias P. Niebel, André R. Studart

Complex Materials, Department of Materials, ETH Zürich, 8093 Zürich, Switzerland

* authors have equally contributed to this work

**Abstract**

Composites are often made heterogeneous in nature to fulfill the functional demands imposed by the environment, but remain difficult to fabricate synthetically due to the lack of adequate and easily accessible processing tools. We report on an additive manufacturing platform to fabricate complex-shaped parts exhibiting bio-inspired heterogeneous microstructures with locally tunable texture, composition and properties and unprecedentedly high volume fractions of inorganic phase (up to 100%). The technology combines an aqueous-based slip casting process with magnetically-directed particle assembly to create programmed microstructural designs using anisotropic stiff platelets in a ceramic, metal or polymer functional matrix. Using quantitative tools to control the casting kinetics and the temporal pattern of the applied magnetic fields, we demonstrate that this robust approach can be exploited to design and fabricate heterogeneous composites with thus far inaccessible microstructures. Proof-of-concept examples include bulk composites with periodic patterns of micro-reinforcement orientation and tooth-like bilayer parts with intricate shapes displaying site-specific composition and texture.

Heterogeneity in composite materials is often encountered in the form of a non-uniform distribution of microstructural features, such as particles, pores, chemical species or living cells, throughout a macroscopic part. Tailored distribution of these features can be an effective means to generate heterogeneous graded materials that combine antagonistic properties or that display functionalities that would not be accessible in uniform microstructures. The core-shell porous architecture of biological bones and engineered sandwiched foams is a typical example of mechanically functional structures that combine opposing properties like stiffness and low weight into a single component (*1*). The emergence of other unusual functionalities through heterogeneous design is also well illustrated, for example in optics, by the siliceous skeleton of marine sponges and by synthetic glass fibers, both of which display a chemical gradient to confine light within their high refractive index core and thus enable efficient transmission using silica glass as single base material (*2*).

While several other examples of heterogeneous architectures exist among man-made materials, living organisms have taken this concept to unprecedented levels by designing heterogeneous composites with exquisite control over the local chemical composition and the texture at multiple length scales (*3-5*). The term texture is referred here to the orientation of building blocks in a particular direction at any given length scale of the material. Such microstructural design results from a long evolutionary process that has gradually crafted the material to satisfy specific demands of the surrounding environment (*6*). Remarkably, some universal structural motifs have emerged from this natural selection process and are conserved in numerous distinct living species across different phyla. These include for instance the prismatic layers of teeth (*7*) and mollusk shells, (*8*) the plywood fiber architectures of fish scales, (*9*) insects, (*10*) crustaceans (*11*) or plants (*12*), the concentric plywood structures around bone osteons (*13*) and in wood cell walls (*14*) and the brick-and-mortar (*15*) and cross-lamellar arrangements (*16*) of mollusk shells. Such patterns can be found at multiple length scales within the structure and often also involve subtle modification of the local composition or the degree of crystallinity (*17*). Natural heterogeneous architectures are created by combining some of these motifs in a single component of well-defined macroscopic shape. A prominent example is the combination of a prismatic external layer with an internal brick-and-mortar, plywood or mesh layer. This architecture is found in natural composites as diverse as teeth, mollusk shells and stomatopod clubs, some of which are produced from genetically distant organisms (*18*). Because they are usually created through the sequential deposition of matter, many of these natural structures retain a distinct multilayered architecture spanning over several length scales.

In contrast to the rich structural diversity observed in nature, heterogeneous design in synthetic materials remains challenging due to the lack of processing routes to achieve the level of multiscale structural control found in biological composites (*4*). To address such limitation, several engineering approaches have been developed in order to enable deliberate tuning of the texture (*19*) and local composition of man-made composites. (*20*) One strategy towards texture control is to use external magnetic fields to program the orientation of anisotropic building blocks in a suspension followed by consolidation of the fluid phase to generate a ceramic or composite (*21-25*). This has led, for example, to polymer-based bilayer composites that can be tuned to display combined strength and wear resistance (*23, 26*) or to exhibit unusual self-shaping effects in response to external stimuli (*27*). Another approach to achieve texture in specific directions is to use ice crystals as templates for the assembly of suspended particles into lamellar structures. The resulting porous scaffolds can then be further densified through subsequent pressing and sintering steps and possibly infiltrated with a second phase (*28, 29*). Such ice templating route has been effectively exploited to fabricate bulk ceramics and composites with outstanding strength, stiffness and crack propagation resistance (*30, 31*). Layer-by-layer assembly routes have also been extensively exploited for the fabrication of nacre-like thin films that combine high in-plane mechanical properties, transparency and flame-retardancy (*32, 33*).

In spite of the exciting developments towards the realization of increasingly intricate bioinspired structures, the creation of composites with complex macroscopic geometries and heterogeneous architectures programmed to fulfill specific functional demands have not yet been demonstrated. Inspired by the layer-by-layer approach used by living cells to fabricate composite materials in nature, we report a novel additive manufacturing approach to create complex-shaped lamellar composites with locally controlled texture and highly mineralized compositions. This is achieved by combining a well-established slip casting process to fabricate complex-shaped components from particle suspensions with a recently developed approach to control the orientation and distribution of anisotropic building blocks using magnetic fields. This simple and robust technology allows for the design and fabrication of bioinspired composites with intricate geometries and heterogeneous architectures that are locally tuned to fulfill site-specific functional demands. Combined with common processing routes for ceramics and composites, novel functional composites with reinforcement volume fraction up to 100 vol% can be reached in a fabrication timescale significantly faster than in both natural and existing synthetic processes.

The fabrication of components through slip casting involves the deposition of a fluid suspension of particles into a dry porous mold of pre-defined geometry and pore dimensions typically smaller than the particle size (fig. 1A). Wetting of the pores of the mold generates capillary forces that continuously remove the liquid phase from the suspension to build a layer of jammed particles next to the mold wall, also known as the cake layer. This continuous assembly process is characterized by a jamming front, which is defined as the boundary between the dispersed particles in the fluid phase and the jammed particles in the cake layer. Besides capillary forces, vacuum can also be used to extract the liquid phase through the walls of the porous mold.

To control the texture of the as-deposited anisotropic particles throughout the thickness of the cake layer, we apply an external magnetic field in a pre-defined direction relative to the mold wall. The anisotropic particles are coated with superparamagnetic iron oxide nanoparticles (SPIONs) to become magnetically responsive, following the protocol established by Erb *et al* (*23*). Changing the direction of the magnetic field as a function of time leads to heterogeneous composites with any desired orientation of anisotropic particles throughout the thickness of the material. In this scheme, the consolidation front works to fix the orientation of anisotropic particles determined by the imposed magnetic field. To define the instantaneous direction of the applied field and thus program the local texture of the composite it is crucial to determine the actual position of the consolidation front as a function of time. This information allows us to establish a simple correlation $t = f(x)$, which defines the time ($t$) required for the consolidation front to reach a given position ($x$) relative to the mold wall (fig 1B). For a target orientation of anisotropic particles ($\phi$) at any position ($x$) along the heterogeneous composite ($\phi = f(x)$), it becomes straightforward to deduce how the angle between the plane of the imposed rotating magnetic field relative to the mold should be changed over time ($\phi = f(t)$) to achieve the desired local texture control (fig. 1C,E). The use of platelets under a rotating external field allows us to describe their orientation by taking solely one angle $\phi$. It is important to note though that this rationale can be extended to more complex systems combining for example 1D reinforcements and static external fields, which require more angular variables to describe the higher degrees of freedom involved.

While magnetic fields enable site-specific texture, the additive nature of the slip casting process allows for compositional variations through the material's thickness by simply changing the constituents suspended in the fluid phase, as schematically depicted in fig. 1D,E. Possible constituents can vary widely in type, morphology, chemical nature and size, from molecules and particles to droplets and bubbles and even living cells. Such building blocks can be used alone or combined to form tunable gradients throughout the thickness of the cast material. Combinations comprising more than one type of magnetically responsive constituent are also possible. This adds one more control parameter to the process, which describes the volume fraction of constituent $i$ ($v_i$) as a function time: $v_i$ = $f(t)$. Similarly to the textural control, such parameter can be deliberately tuned to achieve a target compositional profile ($v = f(x)$) for a given casting kinetics ($t = f(x)$).

For the simple case of anisotropic particles as single suspended constituent, we demonstrate the texture control that can be achieved through magnetically-assisted slip casting (MASC) by establishing quantitative correlations between the functions $t = f(x)$, $\phi = f(t)$ and $\phi = f(x)$ in a thoroughly designed experimental series. In this example, the fluid phase consists of a suspension of magnetically responsive alumina platelets electrosterically dispersed in water, whereas the porous mold is a disc-shaped planar substrate made out of gypsum. In the absence of vacuum, particle deposition at the substrate surface is driven by the capillary pressure $\Delta P$ that develops across the meniscus of water wetting the pores of the mold. Assuming that the growth rate of the cake thickness is controlled by the diffusion of water through the newly formed layer of jammed particles, one can show that the time required for the consolidation front to reach a position $x$ from the mold surface is given by: (*34*)

$$t(x) = Ax^2, \text{ with } A = \frac{\eta R}{2J\Delta P}$$

Eq. 1

where $\eta$ is the viscosity of the liquid, $R$ is the hydrostatic resistance of the cake layer and $J$ is the ratio between the volume of cast layer and of the extracted liquid. The squared dependence of the elapsed time $t$ on the position $x$ was experimentally confirmed by tracking the consolidation front developed during slip casting of a 25 vol% platelet suspension (fig. 2A). For the porous gypsum mold used in this study, we find the constant A from equation 1 to be equal to 3.5 s.mm$^{-2}$.

Control over the orientation of platelets at the jamming front is achieved by exposing the casting system to an external magnetic field of 15 mT rotating at a frequency above 1.6 Hz. Under such conditions, the platelets are expected to biaxially align within the plane of the rotating field (*21-23*). For a single platelet submitted to a lower static magnetic field of 3.5 mT in a Newtonian fluid with a viscosity of 18 mPa.s, alignment in the direction of the applied field should occur in approximately 7 s (*21*). This theoretical estimate is on the same order of magnitude of the experimental value of approximately 3 s observed for the present system under the same conditions.

As an example, the plane of the rotating field with respect to the mold surface is varied between 0 and $\pi$ in a discrete stepwise function of time generating the arbitrary step function shown in fig. 2B. In this case, the dependence of $\phi$ with time *t* is described by the following equation:

$$\phi(t) = \sum_{m=1}^{+\infty} \phi_0 \, H(t - m\tau), \, m \in \mathbb{N} \qquad \text{Eq. (2)}$$

where *H* is the Heaviside step function, $\tau$ is the timescale between steps at a constant pre-defined orientational angle, $\phi_0$ is the desired rotation angle and *m* is an integer. For this function, the period T of the temporal variation of the angle $\phi$ is given by $\pi\tau/\phi_0$.

On the basis of equations 1 and 2, one can derive the following relation for the local texture within the cast material as a function of the position $x$ from the mold wall:

$$\phi(x) = \sum_{m=1}^{\infty} \phi_0 \, H\left(x - \sqrt{\frac{m\tau}{A}}\right), \, m \in \mathbb{N} \qquad \text{Eq. (3)}$$

This implies that the *n*$^{th}$ complete rotation of the platelets from $\phi$ = 0 to $\phi$ = $\pi$ will occur at distances $x_n$ from the surface given by:

$$x_n = \sqrt{m\tau/A} \text{ for } m \text{ values equal to } n\pi/\phi_0. \qquad \text{Eq. (4)}$$

The predictive nature of this relation was experimentally assessed by performing slip casting experiments with suspensions exposed to magnetic fields exhibiting three different step-wise patterns with increasing timescales $\tau$ and a constant angle $\phi_0$ of $\frac{\pi}{4}$, as indicated in fig. 2B. Because of the brown color of the SPIONs, platelets aligned perpendicular to the plane of view generate a darker coloration on the sample surface in comparison to those aligned in a parallel configuration. This alignment dependent color change allows us to readily identify by naked eye the periodic pattern created along the height of materials cast in the presence of the aforementioned step-wise magnetic field profile. The optical microscopy images shown in fig. 2C confirm our ability to create structures with platelets aligned in a well-defined periodic texture using this approach. As expected, the pitch *p* of the observed periodic microstructure is larger in specimens subjected to angular profiles with longer timescales $\tau$. More importantly, we find that the experimentally programmed periodic pattern agrees well with the profile expected from equation 3, confirming the predictive power of our theoretical estimate (fig. 2C). This is evident when the distances $x_n$ given by equation 4 are superimposed on top of the optical microscopy images shown in fig. 2C. As observed, the non-linear growth of the jamming front (fig. 2A) results in a decrease of the pitch as a function of the distance from the mold surface.

For elapsed times significantly longer than the timescale $\tau$ ($t \gg \tau$), one should indeed expect the pitch *p* to vary with the position *x* from the bottom layer according to the following simple equation (for derivation see supporting material (*35*)):

$$p = \frac{\tau}{2Ax} \qquad \text{Eq. (5)}$$

Remarkably, the pitch predicted by equation 5 agrees well with the experimental values obtained for all the three investigated timescales $\tau$ (fig. 2C). The agreement improves for short timescales, for which the initial assumption of $t \gg \tau$ is better fulfilled. For this particular example, the pitch follows an inverse dependence on the position within the cast material. It is important to note though that the pitch and the angular gradient within each period can be deliberately controlled by tuning either the time-dependent orientation of the magnetic field ($\phi(t)$) or the growth rate of the cast layer ($x(t)$). Fig. 1C illustrates how $\phi(t)$ can be programmed to result in a periodic textured architecture of fixed pitch. Alternatively, periodic structures with a constant through-thickness pitch can be easily realized by applying an underpressure beneath the porous mold that linearly increases with time. According to equation 1, this would lead to a constant growth rate of the cast layer, which ultimately translates into a fixed pitch across the structure. As another example, a shell-inspired cross-lamellar structure can be constructed by alternating the alignment angle between $\frac{\pi}{4}$ and $-\frac{\pi}{4}$.

A particular advantage of the proposed processing route compared to other additive manufacturing technologies is that the constituents suspended in the liquid can be fully dispersed at rather low volume fractions in the fluid phase before consolidation at the jamming front. By minimizing steric hindrance effects, this dilute condition enables accurate control over the orientation of anisotropic constituents and leads to the formation of jammed layers with high volume fractions of the dispersed phase. For the alumina platelets studied in this work, we were able to achieve volume fractions as high as 35 vol% in the cast layer as opposed to maximum values of up to 27 vol% obtained with initially concentrated suspensions of platelets of similar aspect ratio (*26*). A high volume fraction of well-aligned platelets after casting is crucial to obtain textured composites with inorganic content that approaches those of highly mineralized biological materials (95 vol%).

The well-aligned layered structures obtained via MASC can be further processed through a sequence of drying, pressing and, optionally, infiltration steps to result in large-scale nacre-like bulk composites with outstanding texture and functional properties (fig. 3A). We show that this simple processing route not only enables reproducible fabrication of nacre-like composites with macroscopic size in the centimeter scale, but also allows for microstructural design through mineral bridge formation combined with compositional control of dispersed and continuous phases over a wide range of volume fractions.

Uniaxial pressing at ambient temperature using pressures up to 100 MPa is used to increase the initial volume fraction of platelets from 35 to up to 60 vol%. To obtain complex shaped parts with such high densities, cold isostatic pressing with pressures approaching 1 GPa can be applied. Further densification is also possible by uniaxially hot pressing the porous structure at temperatures and pressures in the ranges of 1000-1500°C and 0-60 MPa, respectively. The hot pressing step can be done either following cold pressing or as sole densification procedure. This allows us to reach volume fractions spanning from 40 to 100% of inorganic platelets, while keeping a very high degree of alignment throughout centimeter-sized samples (fig. 3B, C). A second continuous phase can be introduced into the lamellar structure either by incorporating particles in the initial casting suspension or by infiltrating the scaffolds after pressing and sintering. This enables the creation of functional composites with unique electrical, thermal and mechanical properties (fig. 3D, E, F). For example, the addition of inorganic silica and alumina nanoparticles in the initial slurry leads to a refractory continuous phase that makes the composite suitable for high-temperature applications. Alternatively, small amounts of magnetized metal flakes can be suspended and co-aligned together with the alumina platelets to create highly textured metal-matrix composites with metal-like electrical and thermal conductivities. Finally, infiltrating porous sintered inorganic scaffolds with monomers and polymerizing them subsequently results in lamellar composites with a large periodic variation in elastic modulus at the microscale, which can be potentially explored in applications requiring simultaneously high stiffness and damping properties.

All these enticing functionalities are combined with outstanding mechanical properties by exploiting the microstructural and compositional design capabilities offered by the proposed processing route. The presence of a significant volume fraction of stiff and strong alumina platelets leads to composites with high fracture strength and elastic modulus with the added benefit of an increasing crack growth resistance (fig. 3F). In the case of the copper-alumina composites, the higher ductility of the metal matrix provides an intrinsic mechanism that works to increase the fracture toughness of the lamellar composite. To provide another extrinsic toughening mechanism in addition to the expected crack deflection at oriented platelets, we also create nacre-inspired mineral bridges between platelets in polymer-based and all-ceramic composites. Such bridges are formed by introducing silica or alumina nanoparticles, respectively, in the initial casting suspension, which are later partially sintered on the surface of opposing adjacent platelets. This combination of toughening mechanisms leads to a remarkable increase in the fracture toughness of the composites, which is clearly captured by the rising crack growth resistance curves shown in fig. 3F. By increasing the fracture toughness of alumina by a factor of 3 to 5 compared to that of an untextured polycrystalline microstructure, these examples demonstrate the major benefit of replicating biological design principles as a means to create functional high-performance synthetic materials. (*36*)

To finally illustrate the potential of the MASC process in combining texture and compositional control with outstanding functional properties in a macroscopic complex shaped part, we fabricate exemplary composites that reproduce the intricate microstructure and geometry of a biological tooth (fig. 4). To this end, a porous gypsum mold is first prepared as an imprint of an actual human molar. A bilayer structure that mimics the dentin-enamel layers of natural tooth was then fabricated by sequentially casting two different aqueous suspensions into the complex-shaped porous mold (fig. 4A). Both suspensions consisted mainly of 20 vol% of alumina platelets and 13 vol% of isotropic silica or alumina nanoparticles suspended in a 5 wt% poly(vinyl pyrrolidone) water solution. The suspension used to mimic the dentin-like layer contained only alumina with average size of 150 nm as isotropic nanoparticles (13 vol%). To obtain a denser and harder material in the outer enamel-like layer, we replaced 4 vol% of such alumina nanoparticles by 100 nm silica particles in the casting suspension (fig. 4B). The silica nanoparticles form a liquid phase during the sintering that increases drastically the bonding strength between platelets (*37*). The casting procedure was carried out in the presence of a rotating magnetic field so as to replicate the orientation of reinforcing elements found in natural teeth. Such reinforcing building blocks are typically aligned perpendicular to the cusp surface within the enamel layer and parallel to the surface in the underlying dentin layer. The as-cast bilayer structures were sintered at 1600 °C for 1 hour and subsequently infiltrated with a mixture of acrylic monomers and initiator conventionally used in dentistry to create a tooth-like complex-shaped composite with 50 to 64 vol% inorganic phase in dentin-like and enamel-like layers, respectively (fig. 4).

Optical microscopy of the resulting macroscopic synthetic tooth reveals the great fidelity achieved by using the MASC process to replicate the finest details of the geometrical contours of the biological counterpart (fig. 4C). A closer view into a specific region of the tooth-like structure using electron diffraction X-ray mapping (EDX) shows the presence of silicon atoms exclusively in the outer enamel-like layer (fig 4D), confirming the suitability of this additive manufacturing approach for the creation of such intricate lamellar structures. Scanning electron microscopy was also performed to assess the orientation of reinforcing platelets in each one of the individual layers. The SEM micrographs reveal that the local orientation of reinforcing platelets successfully matches the targeted texture ($\phi$ = f(*x*)), confirming that the jamming front effectively fixes the instantaneous alignment direction imposed by the external magnetic field during the casting process (fig 4E). A higher local density of the scaffold is also noticeable within the enamel-like layer as compared to the internal layer mimicking dentin. This indicates that the addition of small volume fractions of silica nanoparticles into the composition was successful in promoting higher local densification of the inorganic framework within the outer layer. To evaluate if the distinct site-specific texture and density of the two layers impacts the local mechanical properties of the heterogeneous composite, we measured the microhardness of the material within the boundary between the enamel- and dentin-like layers. Similarly to the dentin-enamel junction (DEJ) of natural tooth, we find that the local hardness is highest at the outermost layer and continuously decreases at the bilayer interface before reaching a lower plateau value within the innermost dentin-like side (fig. 4F). The gradual transition at the synthetic DEJ occurs within length scales on the order of 100 micrometers and is caused by a combined change in platelet orientation and density across the bilayer interface. This is supported by the strong correlation of the local hardness with the silica content and the platelet orientation. Our results indicate that the proposed technology can be an effective approach to generate multilayered composites with tunable mechanical and functional gradients at the microscale.

On the basis of this proof-of-concept example, we demonstrate that MASC is a powerful and fast additive manufacturing route for the fabrication of heterogeneous synthetic composites that are designed to match the functional demands of specific aimed applications. The possibility offered by this technology to design the local texture and composition of complex-shaped composites in highly mineralized systems significantly expands the design space thus far available for the creation of biologically-inspired materials. This provides an exciting opportunity to implement in synthetic composites some of the unique design principles emerged from the evolution of biological materials and to eventually create a wide range of functional bioinspired material systems that are not accessible using state-of-the-art technologies.

**Acknowledgments**
We thank R. Libanori, D. Carnelli, N. Ghielmetti, J. Reuteler, B. Wegmann and P. Kocher

for experimental assistance and discussions. We acknowledge internal funding from ETH Zürich and the Swiss National Science Foundation (grant 200020_146509), as well as support by the Center for Optical and Electron microscopy of ETH Zürich (ScopeM).

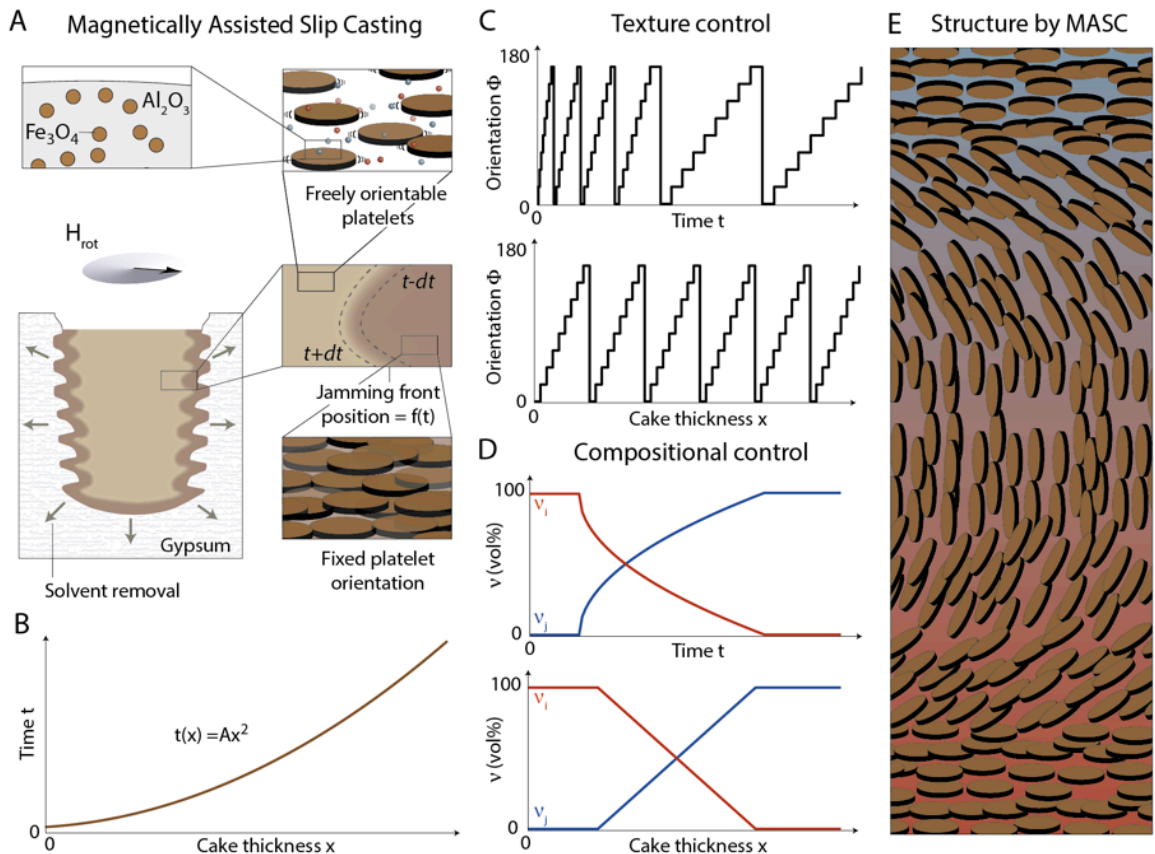

**Fig. 1. Overview of the magnetically-assisted slip casting (MASC) process.** (**A**) Schematics illustrating the casting of an exemplary suspension of SPION-coated alumina platelets into a complex-shaped porous mold in the presence of an external rotating magnetic field. Different additives can be present during the casting, here pictured by two different colored nanoparticles. The dynamic jamming front that enables consolidation of the magnetically-aligned platelets is highlighted. (**B**) Time required for the consolidation front to reach a given position (*x*) relative to the surface of the mold (*t* = f(*x*), eq. 1) for the case of a capillary-driven process. (**C,D**) Programming of (**C**) the orientation of the rotating magnetic field relative to the mold surface and (**D**) the relative fraction of constituents in the suspension as a function of time ($\phi$ = f(*t*) and $v_i$ = f(*t*)) to generate targeted local texture and chemical composition throughout the structure thickness ($\phi$ = f(*x*) and $v$ = f(*x*)). (**E**) The expected variation in local orientation of platelets and chemical composition of the surrounding matrix as a function of distance from the mold surface for the targeted microstructures programmed as shown in (**C,D**).

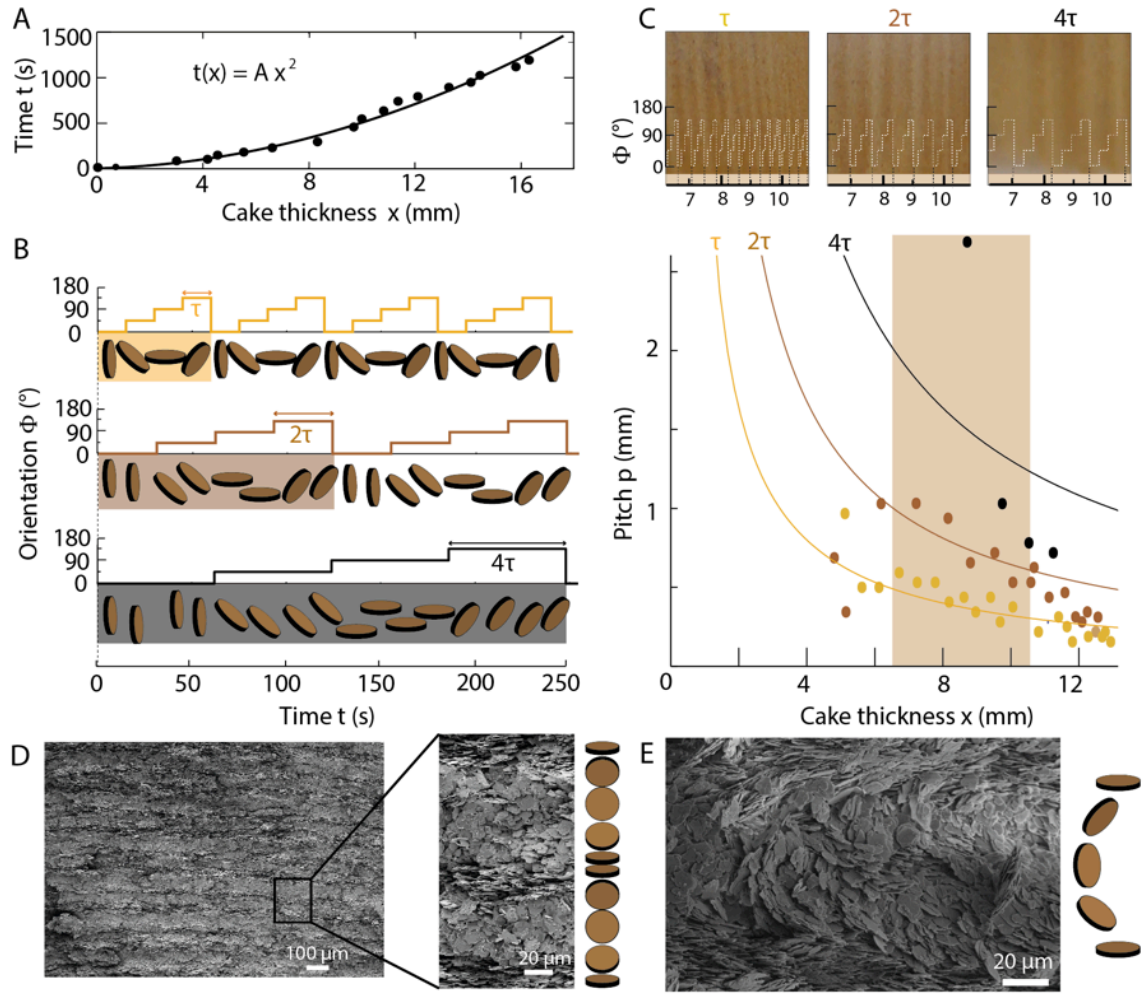

**Fig. 2. Anisotropic structures with periodic platelet orientation patterns obtained through programmed alignment using MASC.** (**A**) Experimental determination of the function $t = f(x)$ (Eq. 1) for a capillary-driven slip casting process, indicating good agreement between theory and experiment. (**B**) Examples of the imposed function $\phi = f(t)$ used to create programmable structures exhibiting platelet orientation patterns with increasing timescales $\tau$. (**C**, top) Optical microscopy images showing agreement between the arrangements of platelets achieved (oscillating color) and the targeted platelet orientation profiles ($\phi = f(x)$, dotted lines). (**C**, bottom) Variation of the pitch ($p$) as a function of the position ($x$) through the thickness of the cast layer for structures with increasing timescales $\tau$. Symbols show experimental data points and lines display theoretical predictions based on eq. 5. (**D, E**) Scanning electron micrographs revealing the periodic platelet orientation pattern that can be generated through MASC. Structures shown in (**C**) were created using 45° angular steps (as depicted in **B**), whereas the architectures displayed in (**D,E**) were obtained using 30 s long 20° angular steps.

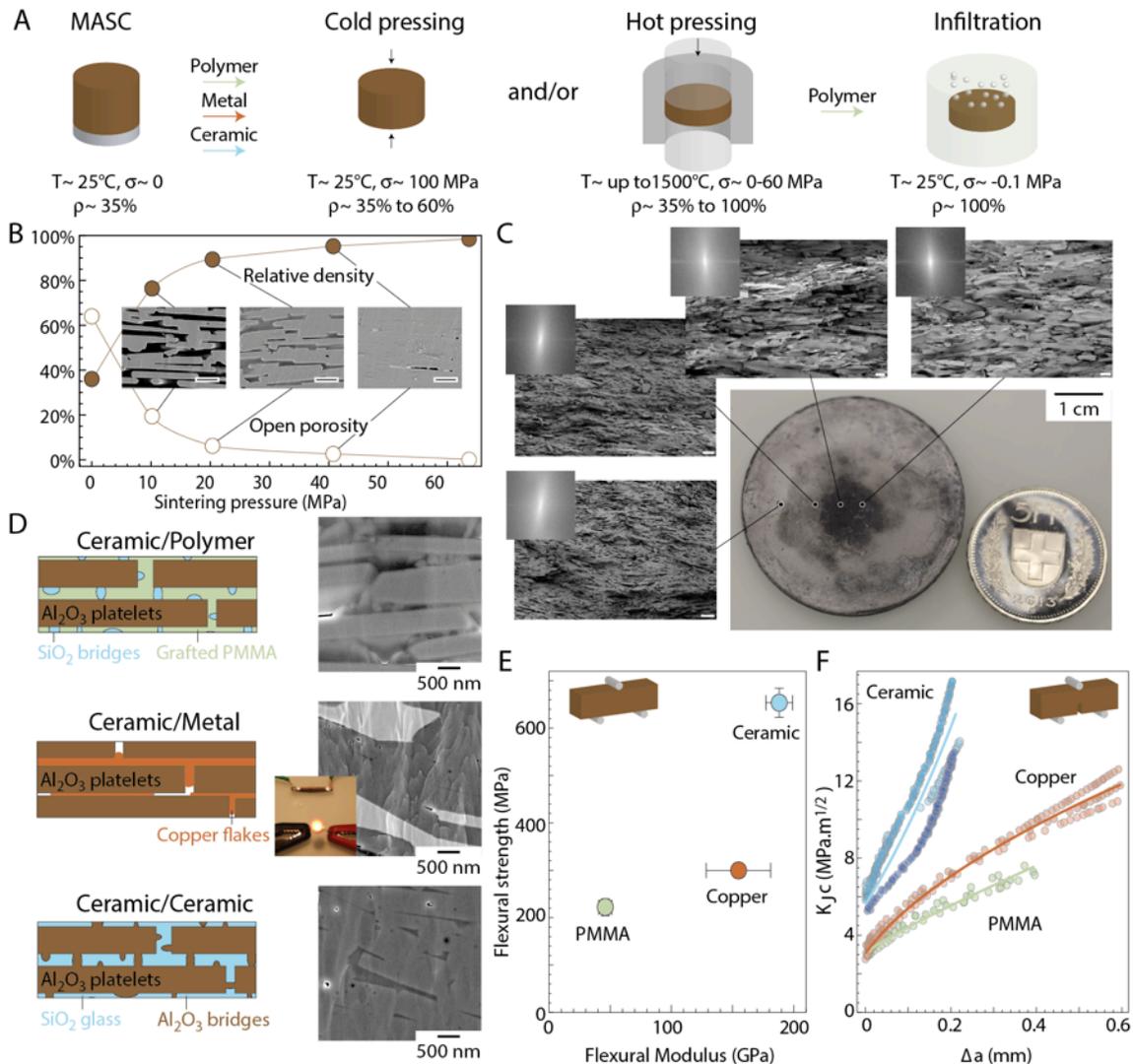

**Fig. 3. Processing of MASC structures into multifunctional composites.** (**A**) Flowchart displaying the pressing and, possibly, infiltration steps that are utilized to tune the volume fraction of anisotropic particles in MASC scaffolds and thus create tunable composites with functional matrices. (**B**) Relative density of MASC scaffolds as a function of the temperature applied in the hot pressing process (dwell time: 30 min). Scale bar: 5 µm. (**C**) Example of bulk part with 5 cm diameter and 3.5 mm thickness exhibiting high degree of platelet alignment throughout the entire specimen (relative density: 93.4 vol% on average). The insets show Fast Fourier Transforms (FFTs) obtained from the SEM images. The highly anisotropic shape of the image in reciprocal space reveals strong alignment of the platelets in the MASC scaffold. Scale bar: 5 µm. (**D**) Examples of nacre-like functional composites exhibiting ceramic, metal or polymer matrices reinforced with 95.5 vol%, 90 vol% and 60 vol% of alumina platelets, respectively. The cartoons and SEM images also display the mineral bridges introduced in composites with polymer and ceramic matrices. (**E**) Flexural mechanical properties and (**F**) crack growth resistance curves of the investigated functional composites.

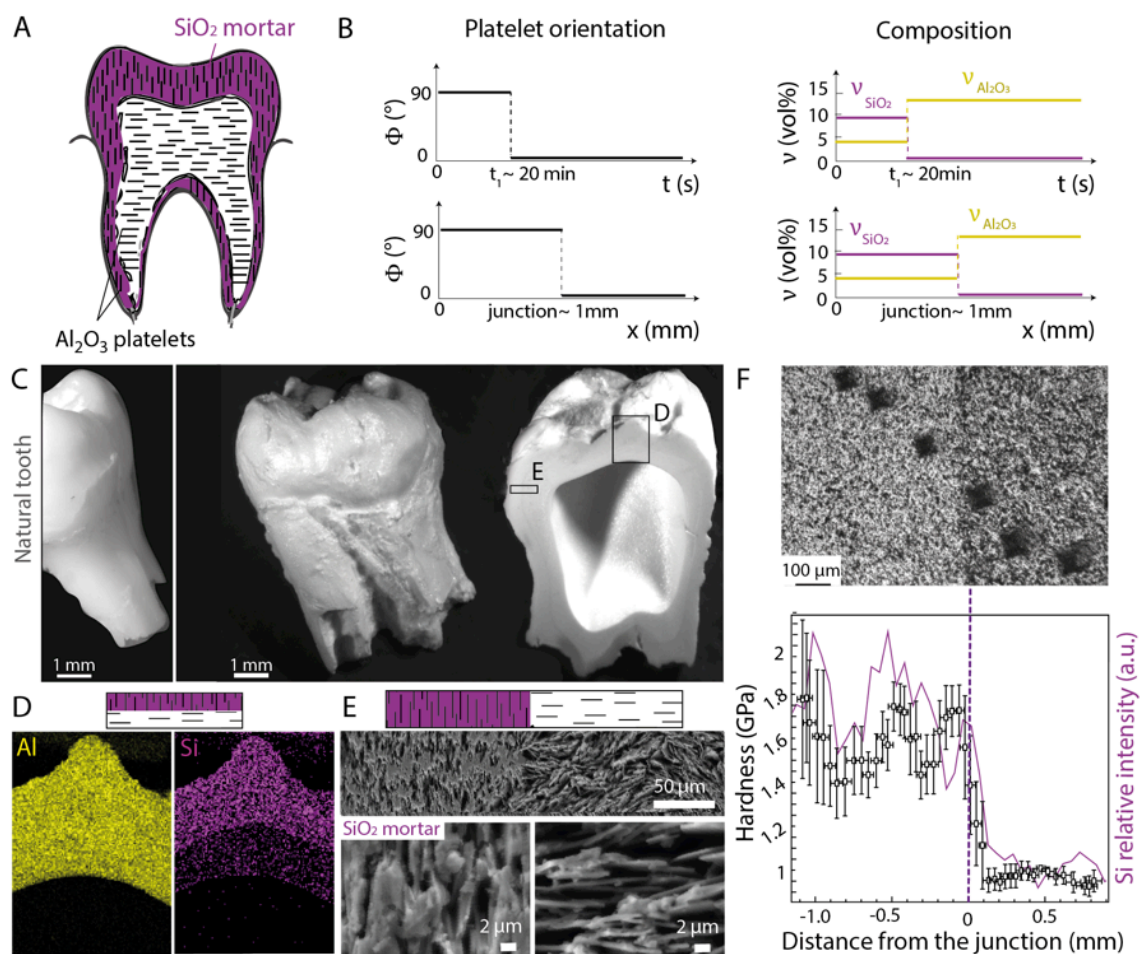

**Fig. 4. Design and fabrication of a bioinspired composite that resembles the complex shape and heterogeneous architecture of natural tooth.** (**A**) Schematic representation of the target complex-shaped part comprising a bilayer with locally distinct platelet orientation and chemical composition. (**B**, top) Programmed temporal pattern of the magnetic field angle ($\phi$ = f($t$)) and of the volume fraction of chemical constituents ($\nu$ = f($t$)). (**B**, bottom) Target local orientation of platelets and chemical composition as a function of the thickness of the cast layer ($\phi$ = f($x$) and $\nu$ = f($x$), respectively). (**C**, left) Natural tooth used as positive template for the fabrication of the complex-shaped porous mold. (**C**, right) Synthetic tooth-like part obtained via MASC, including cross-sectional cut to reveal its internal bilayer structure. (**D**) Elemental analysis of part of the bilayer (window **D** in image **C**), confirming the higher concentration of Si in the outer enamel-like layer. (**E**) SEM image of the synthetic dentin-enamel junction (DEJ), highlighting the distinct platelet orientation in each of the two adjacent layers. (**F**) Microindentation across the synthetic DEJ, indicating the hardness gradient achieved by tuning the local platelet orientation and chemical composition within the composite.

Supporting Online Material for

# Magnetically-Assisted Slip Casting of Bioinspired Heterogeneous Composites

Hortense Le Ferrand, Florian Bouville, Tobias P. Niebel, André R. Studart
Complex Materials, Department of Materials, ETH Zürich, 8093 Zürich, Switzerland

**This PDF file includes:**
- Materials and Methods
- Figs. S1 to S9
- Tables S1 to S5
- References

**Other Supporting Online Material for this manuscript includes the following:**
- Movies S1 and S2

# MATERIALS AND METHODS

## 1 The Magnetically Assisted Slip Casting (MASC) process

### 1.1 Flowchart of the process

The diagram below depicts the main steps involved in the MASC process. Further details are provided in the following sections. A complete list of suppliers of the chemicals used in the experiments is shown in Table I.

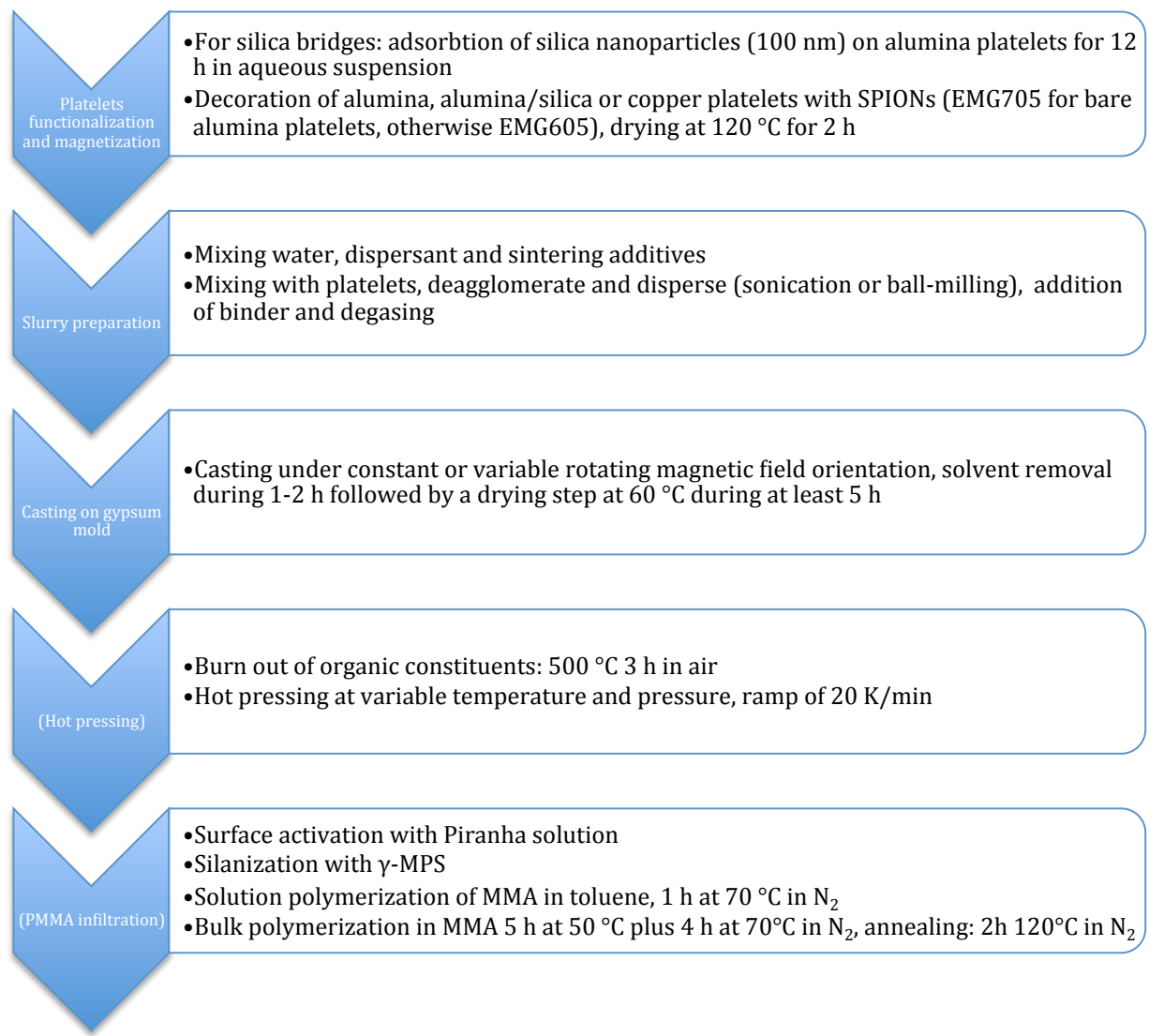

- **Platelets functionalization and magnetization**
  - For silica bridges: adsorbtion of silica nanoparticles (100 nm) on alumina platelets for 12 h in aqueous suspension
  - Decoration of alumina, alumina/silica or copper platelets with SPIONs (EMG705 for bare alumina platelets, otherwise EMG605), drying at 120 °C for 2 h

- **Slurry preparation**
  - Mixing water, dispersant and sintering additives
  - Mixing with platelets, deagglomerate and disperse (sonication or ball-milling), addition of binder and degasing

- **Casting on gypsum mold**
  - Casting under constant or variable rotating magnetic field orientation, solvent removal during 1-2 h followed by a drying step at 60 °C during at least 5 h

- **(Hot pressing)**
  - Burn out of organic constituents: 500 °C 3 h in air
  - Hot pressing at variable temperature and pressure, ramp of 20 K/min

- **(PMMA infiltration)**
  - Surface activation with Piranha solution
  - Silanization with γ-MPS
  - Solution polymerization of MMA in toluene, 1 h at 70 °C in $N_2$
  - Bulk polymerization in MMA 5 h at 50 °C plus 4 h at 70°C in $N_2$, annealing: 2h 120°C in $N_2$

## 1.2 Gypsum mould preparation

Gypsum moulds were prepared according to the supplier's specification using regular tap water at a powder:liquid volume ratio of 2:1. Prior to use, the cast gypsum pieces were dried at least 2 days at room temperature or overnight at 60°C.

## 1.3 MASC experiment

The slurry was cast on the porous mould under a rotating magnetic field. This field allows for the alignment of the magnetized platelets while the solvent is removed from the slurry and the jamming front moves away from the gypsum mould wall. After complete consolidation, the sample was dried for at least 3 h at 60°C. A schematic and the actual set-up used to create horizontally aligned microstructures are shown in Figure S1 and Movie S1.

## 2 Periodic structures made by MASC

## 2.1 Slurry preparation

Platelet magnetization was carried out following the procedure reported by Erb *et al*. (*1*). In a typical experiment, 10 g of alumina platelets were introduced into 300 mL of deionized water and stirred magnetically. 375 µL of ferrofluid EMG705 (Ferrotec, Germany) was pre-diluted in 200 mL water and added to the 300 mL suspension containing the platelets. For complete adsorption of the superparamagnetic iron oxide nanoparticles (SPIONs) on the alumina platelets, the suspension was kept under constant agitation for 24 hours. Complete adsorption was achieved when a clear supernatant was observed after settling of the platelets. The platelets were then filtered and dried at 120 °C for at least 2 h. This procedure can be easily up-scaled to a batch of 100 g of alumina platelets.

A typical slurry composition for MASC consisted of 7.00 g of magnetized platelets added to 5 g of deionized water containing 0.05 g of an ammonium salt of polymethacrylic acid (Dolapix). This results in a suspension with 25 vol% of platelets and 0.35 wt% of dispersant with respect to dry powder mass. Such suspension was deagglomerated by ball-milling with 3 mm alumina milling balls for 3 hours. A ball to powder volume ratio of 1:3 was used in the milling procedure. After ball milling, 0.25 g of the binder poly(vinyl pyrrolidone) (PVP) was added to the slurry to reach a binder concentration of 5 wt % with respect to the water. The slurry was further stirred for an additional hour before thorough degasing in a vacuum of 100 mbar.

## 2.2  Periodic variation of the platelet orientation during MASC

Three transparent poly(methyl methacrylate) (PMMA) tubes were glued on the same piece of gypsum. This mould was then placed 2 cm away from a 400 mT neodymium permanent magnet, which generated a magnetic field strength from 3 to 15 mT in the casting area (movie S2). This magnet was fixed on a motor (motor 1) rotating at a frequency of 1.6 Hz. The rotational speed was controlled by the voltage applied to the motor. Motor 1 was then fixed on another motor (motor 2) whose power supply was commanded by a LabView program (Figures S2 and S3). This configuration enabled the step-wise rotation of motor 2 by applying the following time-dependent voltage:

$$V(t) = \sum_{m=1}^{\infty} V_0 H(t - m\tau), \ m \in \mathbb{N},  \tag{Eq. S1}$$

where $H$ is the Heaviside step function, $\tau$ is the period of time at a constant angle, $V_0$ is the amplitude of the voltage applied to reach the intended angle and $m$ is an integer.

Using this setup during the MASC procedure described in section 1.3, we were able to control the platelet orientation over time. Since this orientation was locally fixed with the jamming front moving forward, this process allowed us to program the local texture of the final architecture.

## 2.3  Cake thickness growth $x(t)$

The kinetics of the jamming front during the slip-casting process was measured on three identical suspensions consolidated onto the same piece of gypsum. To investigate the influence of the platelet orientation on the motion of the jamming front, the suspensions were exposed to rotating magnetic fields applied at 0°, 45° or 90° relative to the surface of the gypsum mould. The cake thickness was accessed by measuring the wet part of an alumina stick dipped in the centre of the casting tube. No major influence of the alignment angle of the platelets on the kinetics was observed, enabling us to extract an average value for the displacement rate of the jamming front in the periodic alignment set-up (Figure S4).

### 2.4 Measurement and prediction of pitch variation with the cake thickness, p(x)

The pitch of each sample from the experimental series with periodically varied magnet orientation was measured after drying at 60°C overnight. The optical images in Figure S5 were analysed using ImageJ and MATLAB to crop regions of interest and adjust color contrast. The intensity profile of each image was obtained by averaging the grey values along the thickness of the layers (grey value = $f(x)$). The minima of these profiles were extracted and the pitch $p(x)$ was calculated as the difference between two consecutive minima.

The variation of the pitch $p$ with the stepping period $\tau$ and with the cast thickness $x$ (Eq. 5 main manuscript) was derived by first defining $p$ as the distance between two layers of same orientation as follows:

$$p(t) = x(t + \tau) - x(t). \tag{Eq. S2}$$

For time periods $\tau \ll t$, one can assume that:

$$x(t + \tau) - x(t) = \tau \cdot \frac{dx}{dt}. \tag{Eq. S3}$$

Since the position of the consolidation front x(t) is equal to $\sqrt{\frac{t}{A}}$ (Eq. 1 in the main manuscript), the derivative dx/dt in equation S3 can be written as:

$$\frac{dx}{dt} = \frac{1}{2Ax} \qquad \text{(Eq. S4)}$$

Replacing this expression in equations S2 and S3, one obtains for $\tau \ll t$:

$$p(x) = \frac{\tau}{2Ax}. \qquad \text{(Eq. S5)}$$

Considering that the cast layer shrinks upon complete removal of the liquid from the cake, the pitch is expected to decrease after the drying process. Following the rationale presented above, one can show that the dried pitch can be obtained through the following relation:

$$p(x) = (1-\delta)^2 \frac{\tau}{2Ax}, \qquad \text{(Eq. S6)}$$

where $\delta$ is the linear shrinkage coefficient after drying the cast sample at 60°C overnight. This coefficient was measured independently on the single-oriented samples used for the determination of the kinetics of consolidation (see section 2.3) and the average value is $\delta$ = 0.237. The derived formula for the pitch (Eq. S6) was used to obtain the fitting curves shown in Figure 2C.

### 2.5 Microstructural characterization

To observe the fracture surface of periodic patterns in a scanning electron microscope (SEM), dried samples were carefully broken to allow for crack deflection along the face of the platelets. After sputter coating with 10 nm platinum, electron microscopy images were taken at an acceleration voltage of 3 kV while tilting the stage at different angles to reveal the microstructure.

## 3 Preparation of nacre-like composites

### 3.1 Preparation of slurry with alumina platelets for compaction study

The compaction behaviour of porous lamellar structures produced by MASC was studied by measuring the density of specimens after sintering at different pressures (Figure 3B, main manuscript). A typical slurry composition for such study consisted of 4.00 g of magnetized platelets mixed with 4.00 g of deionized water containing 0.04 g of Dolapix. This corresponds to 20 vol% of platelets and 0.5 wt% of dispersant with respect to dry powder mass. Sonication was applied during 10 minutes at 60% of its maximal power. Afterwards, 0.202 g of PVP was added to result in a concentration of 5 wt% with respect to water. The slurry was finally degased by applying a vacuum of 100 mbar until bubbles were no longer observed.

### 3.2 Silica adsorption on the platelets, magnetization and slurry preparation

Mineral bridges in nacre-like structures with a polymer matrix (Figure 3D, main manuscript) were created by sintering specimens containing alumina platelets pre-coated with bridge-forming $SiO_2$ nanoparticles. The coating of alumina platelets with $SiO_2$ nanoparticles was carried out following the procedure developed by Libanori *et al*. (*2, 3*). Typically, 10 g of alumina platelets were first added to 400 mL of deionized water. Then, 0.734 g of silica nanoparticles of 100 nm diameter were suspended in 30 mL of deionized water, leading to a slurry that was sonicated for 20 minutes at 60% of the maximum power. The two suspensions were eventually mixed and constantly agitated for at least 12 h to enable complete adsorption of the silica nanoparticles on the surface of the alumina platelets. To magnetize the resulting platelets, 375µL of the ferrofluid EMG 605 were diluted with 70 mL deionized water before they were added to the suspension containing the silica-coated alumina particles. Complete adsorption of the SPIONs on the platelet surface took usually 15-30 min. The platelets were then filtered and dried at 120°C for at least 2 h. The resulting platelets exhibit adsorbed silica nanoparticles (Figure S6, left) with SPIONs visible on top of the silica particles (Figure

S6, right). Assuming complete adsorption of the initially added nanoparticles in the coating procedure, the silica surface coverage was calculated to be 14%.

### 3.3 Metal matrix composites: Milling of copper platelets, magnetization and slurry preparation

Nacre-like composites with a metal matrix were created by introducing magnetized copper flakes into the slurry of alumina platelets used for MASC. In this process, 8.52 g Cu-flakes were first suspended in 50 mL water and planetary milled for 1 h (300 rpm, 1 min interval) with $ZrO_2$ milling media. After separation from the milling media by sieving and washing with water, the Cu-flakes were re-suspended in approximately 0.5 L of water. To magnetize the flakes, 612 µL of the ferrofluid EMG 605 diluted in 25 mL water were added to the suspension of flakes and the resulting mixture was agitated constantly for 5 days. The magnetized Cu-flakes were finally collected by filtration, washed with ethanol and dried under vacuum. (*4*)

To prepare suspensions for MASC, 1.0 g of milled and magnetized copper flakes were mixed with 4.0 g of magnetized platelets in 4.0 g of deionized water containing 0.05 g of Dolapix (20 vol% of alumina platelets and 0.5 wt% of dispersant). The suspension was sonicated for 10 min at 60 % power and 0.2 g PVP was added as binder. The homogeneous slurry was then degassed by applying vacuum of 100 mbar until bubbles were no longer visible.

### 3.4 All-ceramic composites: Nacre-like alumina

A typical slurry composition consisted of 4.00 g of magnetized platelets mixed with 4.03 g of deionized water containing 0.04 g of Dolapix. This resulted in a suspension containing 20 vol% of platelets and 0.5 wt% of dispersant with respect to dry powder mass. Afterwards, 0.068 g of colloidal silica suspension (50 wt% of particles in water) and 0.120 g of alumina nanopowders were added to the suspension with the alumina platelets. Sonication was applied during 10 minutes at 60% of its maximum power. Then, 0.202 g of PVP was added leading to a binder concentration of 5 wt% with respect to water. The slurry was degased by applying a vacuum of 100 mbar until complete removal of visible bubbles.

## 3.5 Casting, cold pressing and hot pressing

The MASC process was performed as described in section 1.3 using PMMA cylinders glued on gypsum substrates used as porous moulds. The PMMA cylinder had the same diameter as the hot pressing die used later in the hot pressing step (20 mm diameter). The sample rotation speed was kept higher than 1 Hz to be above the critical alignment frequency established by Erb *et al* (*1, 5, 6*).

Before uniaxial hot pressing, the binder present in the MASCed sample was removed by a heat treatment at 500°C for 3 h. Then, the samples were placed in a 20 mm diameter graphite die coated with Boron Nitride spray. The heating rate used in the hot-pressing operation was 20°C/min for all the samples. The pressure was applied once the temperature reached 700°C at a loading rate of 4 kN/min. The composition and pressing conditions are described in Table S5.

Some samples were first pressed at ambient temperature and sintered afterwards. In such cases, the uniaxial cold pressing step was carried out in a 30 mm pressing die using a maximum pressure of 100 MPa. The samples were then pressureless sintered at 1600°C for 1 hour to remove the binder and promote bridge formation between the platelets.

## 3.6 Polymer matrix composites: Infiltration of lamellar structure with acrylic monomers

The infiltration of acrylic monomers into the porous lamellar structures fabricated through MASC and their subsequent polymerization was based on a procedure developed earlier for nacre-like composites generated by ice templating.(*7, 8*) To reactivate the surface of alumina platelets coated with $SiO_2$ particles, the sintered porous scaffolds were placed for 30 min in a piranha solution comprising of an equivolumetric mixture of hydrogen peroxide and sulphuric acid. Mild vacuum was applied during this process to ensure complete infiltration of the solution. After the piranha treatment, the scaffolds were washed extensively with deionised water and once with acetone. The samples were soaked over night in acetone and allowed to air dry during 15 min before the next processing step.

To enable strong covalent bonding between the polymer matrix and the inorganic scaffold, the activated surfaces were grafted with acrylate groups using silane chemistry. To this end, the scaffolds were immersed in an equivolumetric solution of acetone and (3-trimethoxysilyl) propylmethacrylate ($\gamma$-MPS). Mild vacuum was applied to ensure complete infiltration. After 10.5 h, the samples were removed from the silanization solution, rinsed with acetone and dried in air.

The polymerization process was conducted in two steps in order to allow for effective adhesion of the polymer phase to the ceramic scaffold. In this protocol, a relatively high initiator concentration is first used which increases the probability of PMMA chains being grafted onto the scaffold surface. These chains can then entangle effectively with the polymer that is synthesized in the subsequent bulk polymerization step. For the first step, the silanized scaffolds were immersed in a solution of 2 volume parts of toluene and 1 volume part of purified methylmethacrylate (MMA) containing 1 wt% of the initiator Azo bis-isobutyronitrile (AIBN) relative to MMA. This solution was deoxygenized by bubbling $N_2$ gas for 20 minutes through a needle penetrating the septum used to cap the poly(propylene) (PP) centrifuge tube used as reaction vessel. To ensure complete infiltration, vacuum of 0.1 mbar was applied for 5 min. The reaction vessel was then flushed once again with $N_2$ gas and placed into an oil bath at 70 °C for 1 h. Afterwards, the reaction solution was removed using a syringe and the scaffold was dried in a vacuum of up to $9.1 \times 10^{-2}$ mbar for 15 min.

In the second step of the infiltration process, degassed and purified MMA containing 0.5 wt% AIBN was added into the reaction vessel so that the scaffolds were completely immersed. Vacuum was applied to favor complete infiltration. Bulk polymerization was conducted by heating the reaction mixture in an oil bath at 50°C for 5 hours and then at 70°C for 4 hours. To ensure complete curing and to remove residual stresses from the polymerization process, the samples were kept in a $N_2$ filled oven at 150 °C for 2 h before being cooled down slowly (~ 30 K/h) below the glass transition temperature ($T_g$) of PMMA (105 °C).

### 3.7 Mechanical testing of notched and un-notched nacre-like composites

The 20 mm diameter disk-shaped samples obtained after sintering, and possibly polymer infiltration, were cut into beams of approximate dimensions 14x2x2 $mm^3$. The un-notched beams for three-point bending tests were then mirror polished and beveled to avoid any crack initiation from the sides.

Single-edge notched beam (SENB) test specimens were first notched with a diamond saw of 300 µm thickness. The bottom of each notch was then sharpened by repeatedly passing a razor blade coated with diamond paste (1 µm). Using this method, the final notch radii were always below 40 µm. At least 3 specimens were tested for each composition and setup. The fracture toughness of nacre-like composites was determined by monotonically loading the specimens to failure at a constant displacement rate of 1 µm.s$^{-1}$. The beam deflection was measured using a linear variable differential transformer (LVDT) setup. Typical load displacement curves from SENB tests for each series of nacre-like composites are shown in Figure S7.

### 3.8 Determination of the crack length

The indirect method often used to determine the length of the crack propagating from the notch is based on the change in the compliance of the SENB specimen during cyclic loading. However in our case this method has proven to be unsuitable because the repeated cycles applied for compliance measurements induce small crack propagation even at low stresses. Instead, we used the simple correlation between compliance and crack length in a SENB test (*9*). The compliance *C* of the beam was obtained from the relation *C = u/f*, where *u* and *f* are the total displacement and force at each point after crack initiation, respectively. From the obtained compliance data, the crack length *a$_n$* at a given step *n* was recursively calculated from the length *a$_{n-1}$* at the previous step using the equation:

$$a_n = a_{n-1} + \frac{W - a_{n-1}}{2} \frac{C_n - C_{n-1}}{C_n} \qquad \text{(Eq. S7)}$$

where *W* is the thickness of the specimen, and *C$_n$* and *C$_{n-1}$* are the compliance values calculated at the *n* and *n-1* steps, respectively.

### 3.9 Details on the J-Integral calculation

To assess the amount of energy dissipated during the propagation of stable cracks through nacre-like composites (crack length < 400 µm), we calculated the J-integral as a function of crack extension, following the procedure previously reported in the literature for the characterization of bone and other heterogeneous materials (*10*). The J-integral comprises an elastic and a plastic contribution. The elastic contribution $J^{el}$ was obtained from the following simple relation derived from linear-elastic fracture mechanics: $J^{el} = K_{IC}^2/E'$, where $K_{IC}$ is the stress intensity factor and $E'$ is the elastic modulus of the material in plane strain condition (e.g. $E' = E/(1 - v^2)$). The plastic component $J^{pl}$ was calculated with the following equation:

$$J^{pl} = \frac{1.9 A^{pl}}{Bb}, \quad \text{(Eq. S8)}$$

where $A^{pl}$ represents the plastic area under the load-displacement curve, $B$ is the specimen lateral dimension and $b$ is the uncracked ligament. To take into account that the ligament $b$ continuously decreases during crack propagation, the following incremental definition of Eq. S8 was used to obtain the plastic contribution of $J$:

$$J_n^{pl} = \left[ J_{n-1}^{pl} + \left(\frac{1.9}{b_{n-1}}\right)\left(\frac{A_n^{pl} - A_{n-1}^{pl}}{B}\right)\right]\left[1 - \frac{a_n - a_{n-1}}{b_{n-1}}\right] \quad \text{(Eq. S9)}$$

An equivalent stress intensity factor $K_{JC}$ that accounts for elastic and inelastic toughening mechanisms during fracture was also calculated as follows:

$$K_{JC} = \sqrt{(J^{el} + J^{pl})E} . \quad \text{(Eq. S10)}$$

The Young's modulus of alumina of 400 GPa was used as the $E$ value for the all-ceramic nacre-like composites. For the metal matrix and polymer matrix composites, a rule of mixture was used to estimate the elastic moduli. This resulted in $E$ values of 372 GPa and 212 GPa for the copper-alumina and the PMMA-silica-alumina composites, respectively.

### 3.10 Electrical conductivity of the nacre-like copper-alumina composite

The electrical conductivity of the copper-alumina nacre-like composites was tested with a standard multimeter and 2 pointed probes. We demonstrate that the electrical conductivity obtained in this composite was sufficiently high to enable lighting of a LED using a 1.7 V laboratory power source (Figure 3D, main manuscript). The wiring was connected to the sample using silver paste.

### 3.11 Microstructural characterization

A composite from each experimental series was cut and polished with Broad Ion Beam milling using the microscopy facilities available at ETH Zürich (ScopeM). Milling was performed using an argon gun accelerated under 6 kV while the sample was wobbled at a middle speed (C3) to avoid heating. Electron microscopy images were taken after deposition of 5 nm layer of platinum on the cut and polished surface.

## 4 Tooth-like synthetic composite made by MASC

A bilayer tooth-like composite was prepared through the sequential casting of two different suspensions of magnetized alumina platelets into a porous gypsum mould. Casting was carried out in the presence of an external magnetic field oriented at distinct orthogonal directions in order to reproduce the local texture of the enamel and dentin layers of natural teeth (Figure 4, main manuscript).

### 4.1 Slurry preparation

Magnetically responsive alumina platelets were prepared as described in section 2.1. For the enamel-like slurry, a suspension of 0.36 g alumina nanoparticles in 2 g water and 0.05 g Dolapix was ball-milled for 24h prior to the addition of 2 g of magnetized platelets. After adding 0.22 g of PVP into the suspension, ball-milling proceeded for another 3 hours. The mixture was thoroughly degassed in vacuum before casting. The slurry for the dentin-like layer was prepared following the same procedure, but using a mixture of 0.09 g of alumina nanoparticles and 0.2 g silica nanoparticles instead of just alumina nanoparticles.

### 4.2 Gypsum mould and original tooth

To obtain the porous mould for MASC, a gypsum slurry was prepared as explained in section 1.2 and casted around a natural human molar. After setting of the gypsum, the resulting porous mould was dried at 60°C overnight before carefully removing the tooth using a dental tool. The negative impression left by the natural tooth in the porous gypsum was then used to prepare the complex-shaped synthetic counterpart (Figure S8).

### 4.3 Magnetically-assisted casting of the synthetic tooth replica

The slurry with the composition for the outer layer of the synthetic tooth was cast first under a rotating magnetic field generated from a 400 mT permanent magnet turning at a frequency above 1.6 Hz. In this first casting, the magnet was oriented such that a vertical alignment of platelets was achieved on the top surface of the tooth. After keeping these conditions for 20 min, this first layer was entirely jammed and the second slurry was poured on top. This time the magnet orientation was turned by 90° to promote platelet alignment in a perpendicular direction with respect to the first layer. After complete removal of the fluid through the pores of the mould, the cast sample was dried at 60°C overnight and then carefully demoulded. The binder was burned out at 500°C for 3 hours before 1h sintering at 1600°C.

### 4.4 Density measurements

Single bulk structures of each layer composition were cast separately in the same condition as in the synthetic tooth. After drying and sintering, the density of each layer was measured using the Archimedes method in water at 23°C and averaged over 3 samples. The open porosity (OP) was calculated using the equation:

$$OP = \frac{m_{la} - m_a}{m_{la} - m_{ll}} * 100, \qquad \text{(Eq. S11)}$$

where $m_a$ is the dry weight of the sample, $m_{ll}$ is the weight of the water-soaked scaffold measured in water, and $m_{la}$ is the weight of the water-soaked scaffold measured in air.

## 4.5 Characterization of the texture and composition

The sintered synthetic tooth was mirror-polished and coated with a layer of 5 nm of platinum before SEM investigation under 10kV acceleration voltage. EDX maps were taken under an acceleration voltage of 20kV.

## 4.6 Infiltration of the tooth replica with dental monomers

The tooth replica was infiltrated with a monomer mixture that is widely used in dentistry. To this end, the porous synthetic tooth was placed in a mixture of the acrylic monomers BisGMA and TEGDMA in equal weight ratios and 0.5 wt% AIBN relative to the total monomer weight. To ensure complete impregnation, vacuum was applied for 5 hours with a minimum pressure of 0.01 mbar. The reaction vessel was then flushed with $N_2$ and warmed in an oil bath at 50°C to initiate the polymerization. After 1 hour, the temperature was raised to 60°C for one additional hour. To ensure complete polymerization the sample was finally annealed at 80°C for 1 hour.

## 4.7 Hardness measurements

Vickers Hardness was measured on the synthetic tooth prior and after the infiltration. Indents were made along a line across the interface between the enamel-like and dentin-like layers using an indentation force of 0.9807 N for the not-infiltrated tooth and of 4.903 N for the infiltrated tooth. The indents were made every 10-40 µm from the outer to the inner layer of the tooth sample. The values were then averaged over 5 measurements within a 100 µm region. Optical images using a microscope operating in the reflective mode were used to localize precisely the position of the indents. The size and shape of the indents were further confirmed using electron microscopy and show the role of the silica glass and polymer phase as mortar between the platelets (Figure S9).

**FIGURES**

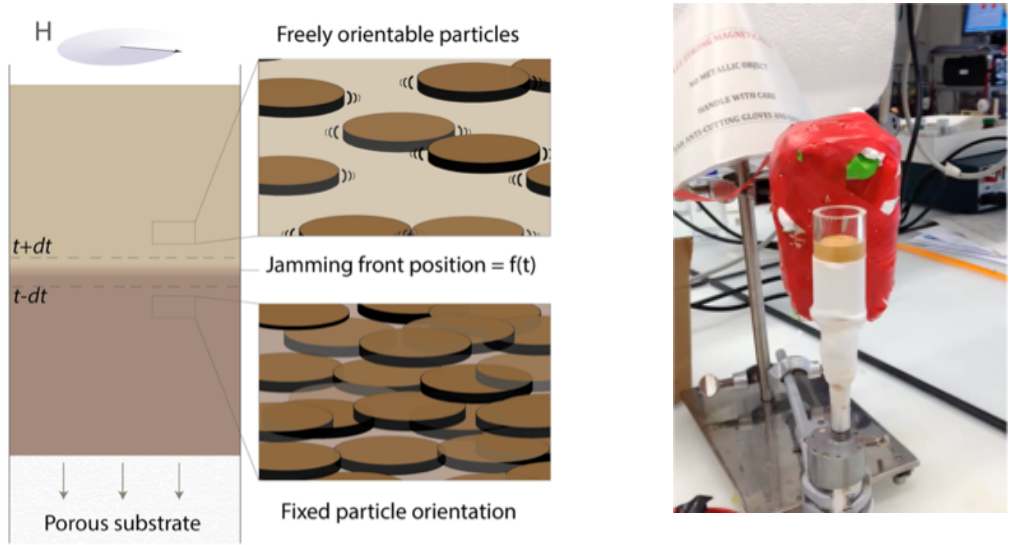

**Figure S1.** Left: Cartoon of the MASC process showing the jamming front and the rotating magnetic field orientation. Right: Actual set-up showing the 500 mT neodymium magnet (in red), the gypsum mould and the transparent PMMA tube containing the slurry. The rotating magnetic field was created by turning the sample in front of the magnet (see also Movie S1**)**

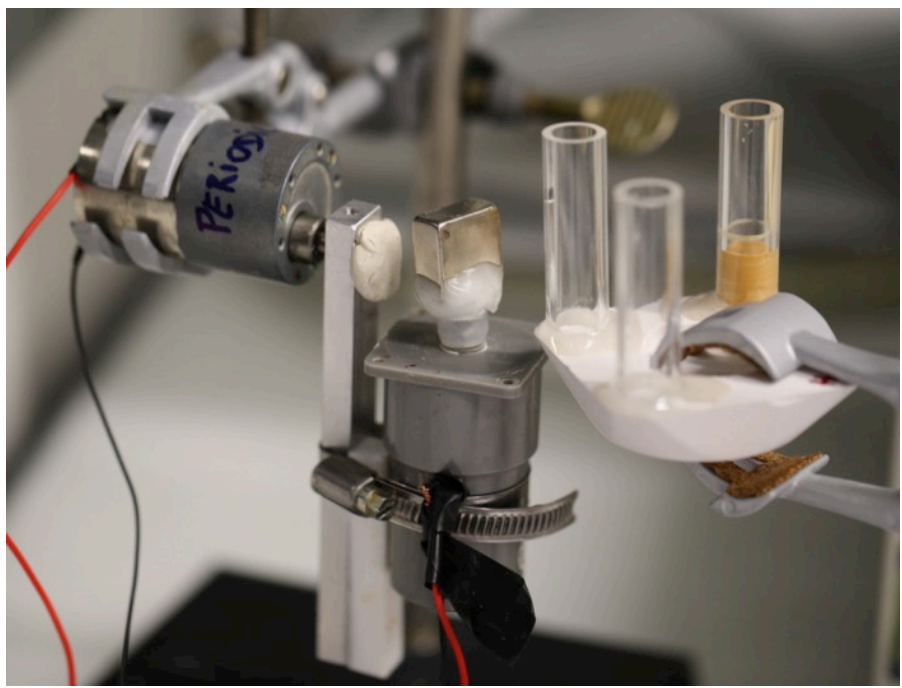

**Figure S2**. Picture of the set up used to apply a stepped magnetic field with different time periods $\tau$ at each angular step (Figure 2B, main manuscript and movie S2). Left: the motor 2 controlled by the LabView program changes the angle of the applied magnetic field with respect to the mould surface. Middle: the 400 mT permanent neodymium magnet attached to motor 1 rotates at 1.6 Hz to align the platelets in 2D parallel to the current rotation plane of the field. Right: the porous gypsum with PMMA tubes. The image shows one slurry (in brown) that is already cast and jammed in one of the PMMA tubes.

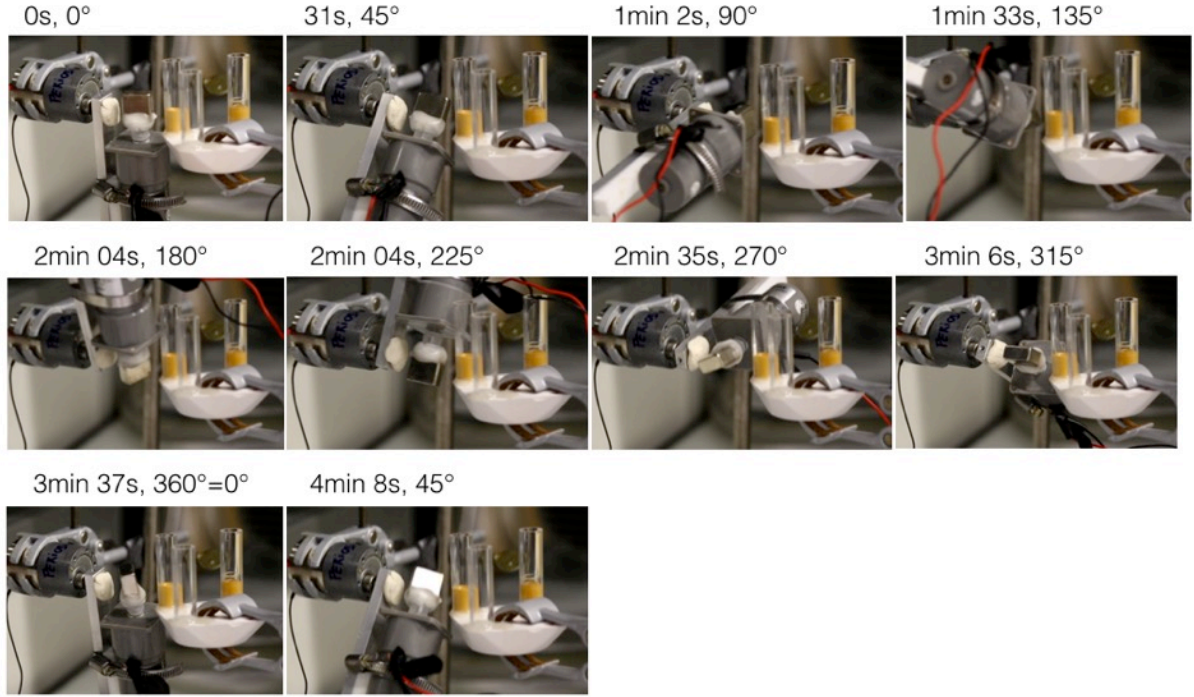

**Figure S3**. Snapshots depicting one full rotation of the motor 2 for the case of $\tau$ =31 s and an angle step of 45° (see also Movie S2).

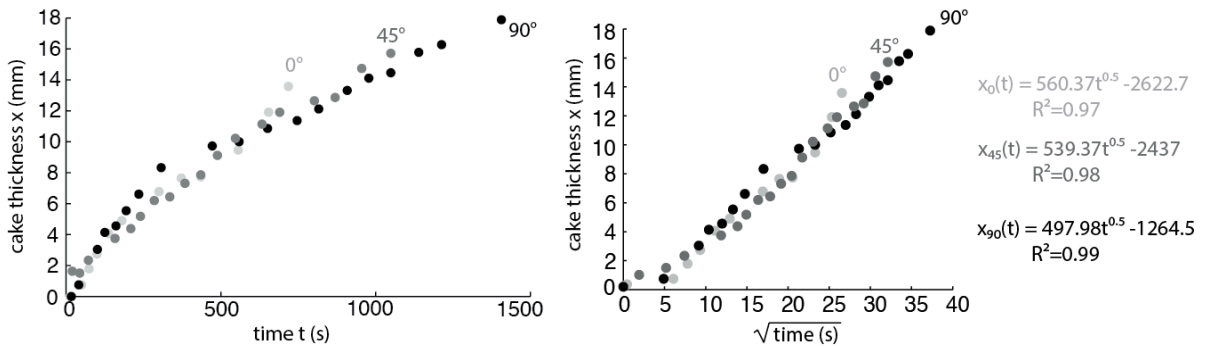

**Figure S4**. Determination of the consolidation kinetics for a platelet suspension cast on a flat gypsum mould while exposed to a magnet aligned at different angles with respect to the mould surface. Left: raw thickness data obtained as a function of time ($t$) for different angles. Right: thickness values plotted as a function of $t^{1/2}$, confirming the theoretical dependence expected from equation 1 (main manuscript).

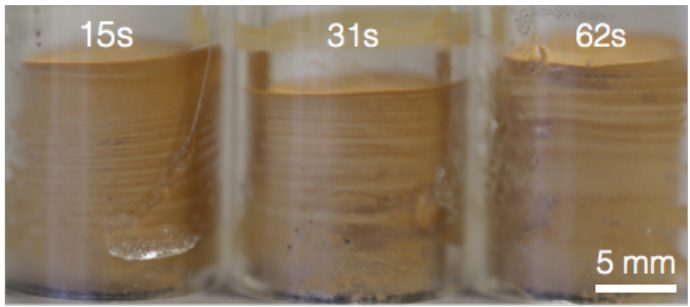

**Figure S5**. Pictures of cast samples obtained from the experimental series with stepped magnet orientations. The step angle was π/4 for each sample and the period increases from 15 s to 31 s to 62 s.

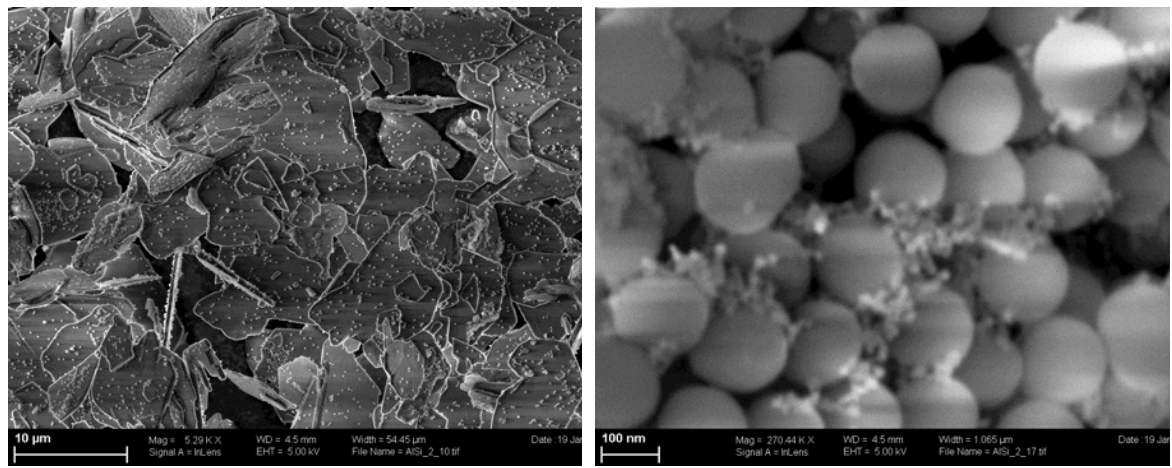

**Figure S6**. Electron microscopy images showing (left) alumina platelets coated with 100 nm silica nanoparticles and (right) a detailed view of the 100 nm silica nanoparticles adsorbed on the platelets to reveal the presence of 12 nm iron oxide nanoparticles (SPIONs) adsorbed onto the silica.

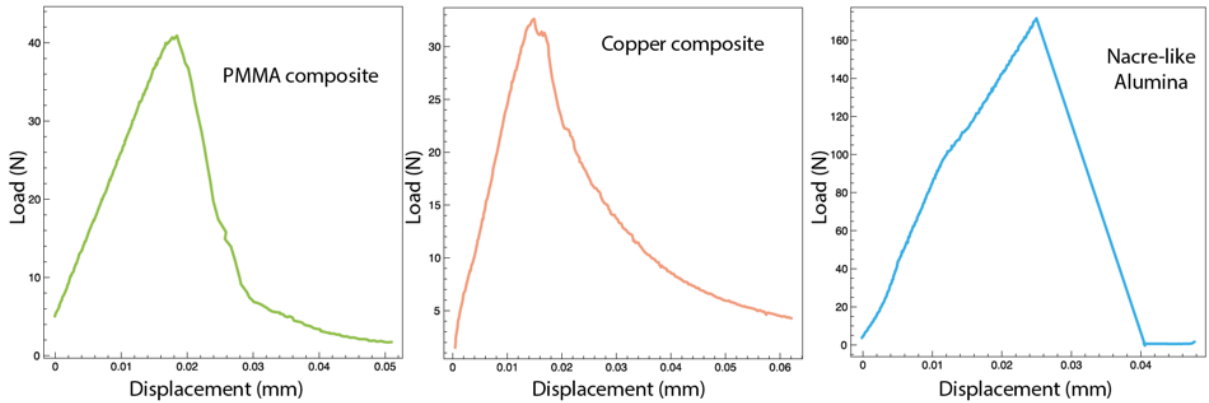

**Figure S7**. Typical load versus displacement curves obtained for the three different nacre-like composites studied in this work. Note that the sample dimensions differ for each type of composites.

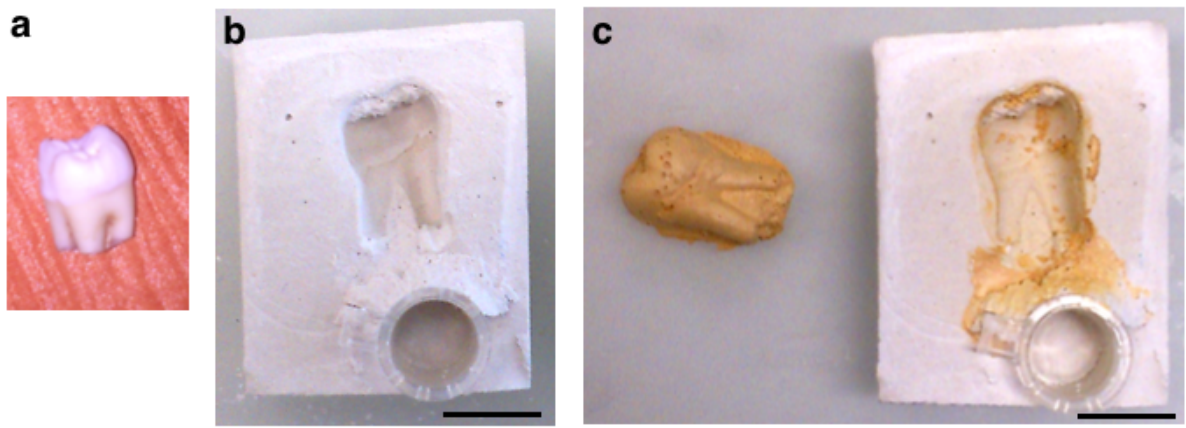

**Figure S8**. Negative impression of (a) a human tooth in (b) a porous gypsum mold. (c) Macroscopic tooth-like structure with fine topographic details obtained through the sequential casting of two suspensions into the negative impression of the porous mould.

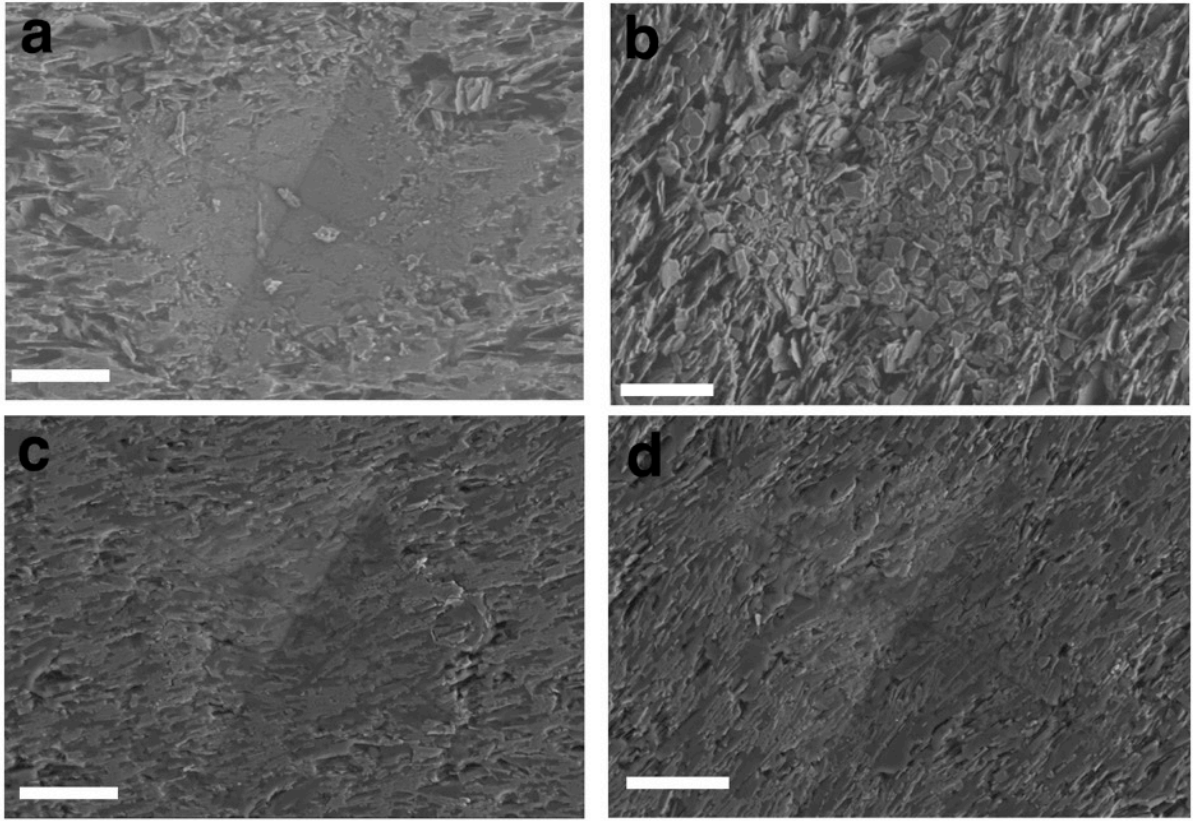

**Figure S9**. Shape of the indents made on the synthetic tooth before (a and b, at 0.9807 N) and after infiltration with the dental monomers (c and d, at 4.903 N). Micrographs (a,c) show the enamel-like layer, whereas images (b,d) display the dentine-like layer. Scale bars: 20 µm.

# TABLES

Table S1. List of raw materials and chemicals

| Material description | Brandname | Provider | Additional information |
|---|---|---|---|
| **Acetone** | Acetone | Sigma-Aldrich (Switzerland) | Reagent grade |
| **Activated alumina** | EcoChromTM MP Alumina B, Act I | MP Biomedicals (Eschwege, Germany) | Used as received |
| **AIBN** | Azo bis isobutyro nitrile 98% | Aldrich (Switzerland) | Used as received |
| **Alumina nanoparticles** | TM-DAR | Taimei (Japan) | Mean diameter 180 nm |
| **Alumina platelets** | Ronaflair WhiteSapphir | Merck (Germany) | Mean diameter 10µm, thickness 250nm, specific surface area 1.5 m2/g, |
| **BisGMA** | Bisphenol A glycerolate dimethacrylate | Aldrich (Switzerland) | Used as received |
| **Boron Nitride Spray** | | Molyduval | |
| **Colloidal silica suspension** | Ludox 50M | Sigma-Aldrich (Switzerland) | Mean diameter 40 nm |
| **Cu-flakes** | Premium copper flakes C125 | Metal flake technologies LLC (San Diego, USA) | Size ~ 125 µm, after milling ~10-30 µm |
| **Dolapix** | Dolapix CE 64 | Zschimmer & Schwarz (Lahnstein / Germany) | Polymeric dispersant based on acrylic acid |
| **Ethanol** | Ethanol | Fluka (Switzerland) | Reagent grade |
| **Ferrofluid** | EMG 605 | Ferrotech (Bedford, USA) | Cationic surfactant |
| **Ferrofluid** | EMG 705 | Ferrotech (Bedford, USA) | Anionic surfactant |
| **Hydrogen Peroxide** | Hydrogen Peroxide ≥ 35% | Sigma-Aldrich (Switzerland) | Used as received |
| **MMA** | Methylmethacrylate 99%, stabilized | Acros (Belgium) | Purified by passing over a plug of activated alumina |
| **Modelism plaster** | | Boesner | |
| **PVP** | Polyvinylpyrrolidone | Sigma-Aldrich (Switzerland) | Mw~360 000 g/mol |
| **Silica nanoparticles** | Silica monodisperse nanosphere | AngstromSphere (Bedford, USA) | 100 nm diameter |
| **Sulphuric acid** | Sulphuric acid 96-97% | Sigma-Aldrich (Switzerland) | Used as received |
| **TEGDMA** | Tri ethylene glycol dimethacrylate | Aldrich (Switzerland) | Used as received |
| **Toluene** | Toluene | Sigma-Aldrich | Reagent grade |

|  |  | (Switzerland) |  |
|---|---|---|---|
| γ-MPS | (3-Trimethoxysilyl) propylmethacrylate 98% | Aldrich (Switzerland) | Used as received |

Table S2. List of equipment used

| Equipment | Name | Provider | Additional information |
|---|---|---|---|
| **BIB** | IM4000 | Hitachi (Japan) | |
| **Hot press** | HP | FCT Systeme GmbH | |
| **Macroscope** | Leica Z16 APO | Leica (Switzerland) | Camera Leica DFC 365 FX |
| **Mechanical testing frame** | Instron 6582 | Instron | |
| **Microindenter** | MXT-a | Wolpert (Germany) | |
| **Microscope** | Leica DM 6000 B | Leica (Switzerland) | Camera Leica DFC 360 FX |
| **Motors** | Modelcraft | Conrad | 174 rpm 12V / 26 rpm 12V |
| **Permanent magnet** | | Supermagnete.ch | Neodymium |
| **Planetary miller** | | Reitsch | |
| **Pressing die, diameter 30mm** | | Eurolabo | |
| **SEM and EDX** | Gemini | Zeiss (Germany) | |
| **Sonicator** | UP200s | Dr. Hielscher | 200 Watts, 24kHz |

Table S3. List of softwares used.

| Software | Provider | Additional information |
|---|---|---|
| **LabView** | | |
| **Image J** | | Image J 1.47v |
| **MATLAB** | Mathworks (USA) | Version 2013 |
| **EDX Software** | ThermoFischer | |

Table S4. Processing conditions and composition of the hot pressed composites. The volume fractions of each phase are given in parentheses.

| Composition and name | Hot pressing temperature/dwell time | Hot pressing applied stress | Final density |
|---|---|---|---|
| **Alumina platelets** | 1400°C/30min | 10 MPa | 76 % |
| **Alumina platelets** | 1400°C/30min | 20 MPa | 89 % |
| **Alumina platelets** | 1400°C/30min | 40 MPa | 95 % |
| **Alumina platelets** | 1400°C/30min | 60 MPa | 98 % |
| **Alumina:Silica (86 : 14) for PMMA composites** | 950°C/60min | 60 MPa | 60 % |
| **Alumina : copper (90 : 10) composites** | 1150°C/60min | 60 MPa | 90% |
| **Nacre-like alumina Alumina : Alumina nanoparticles : silica (95.5 : 3 : 1.5)** | 1500°C/ 30 min | 60 MPa | 98% |

**MOVIES:**

Movie S1 shows the experimental set-up used to align the magnetized platelets into a horizontally textured nacre-like structure using the MASC process.

Movie S2 shows the experimental set-up used to create the periodic structure with 45° stepping angle and 31 s stepping time. The movie is sped up by a factor of 8.